\documentclass[prl,a4paper,reprint,superscriptaddress,nofootinbib,preprintnumbers]{revtex4-1}
\pdfoutput=1

\usepackage{mymacros}

\usepackage{color}
\definecolor{change}{rgb}{0.0, 0.0, 0.0} 

\begin{document}

\title{Path integral optimization as circuit complexity}

\author{Hugo A. Camargo} \email{hugo.camargo@aei.mpg.de}
\affiliation{Max  Planck  Institute  for  Gravitational  Physics (Albert Einstein Institute),\\Am M\"uhlenberg 1, 14476 Potsdam-Golm, Germany}
\affiliation{Department of Physics, Freie Universit\"{a}t Berlin, Arnimallee 14, 14195 Berlin, Germany}

\author{Michal P.\ Heller} \email{michal.p.heller@aei.mpg.de}
\altaffiliation[\emph{On leave of absence from:}]{ National Centre for Nuclear Research, Pasteura 7, 02-093 Warsaw, Poland.}
\affiliation{Max  Planck  Institute  for  Gravitational  Physics (Albert Einstein Institute),\\Am M\"uhlenberg 1, 14476 Potsdam-Golm, Germany}

\author{Ro Jefferson} \email{rjefferson@aei.mpg.de}
\affiliation{Max  Planck  Institute  for  Gravitational  Physics (Albert Einstein Institute),\\Am M\"uhlenberg 1, 14476 Potsdam-Golm, Germany}

\author{Johannes Knaute} \email{johannes.knaute@aei.mpg.de}
\affiliation{Max  Planck  Institute  for  Gravitational  Physics (Albert Einstein Institute),\\Am M\"uhlenberg 1, 14476 Potsdam-Golm, Germany}
\affiliation{Department of Physics, Freie Universit\"{a}t Berlin, Arnimallee 14, 14195 Berlin, Germany}


\begin{abstract}
Early efforts to understand complexity in field theory have primarily employed a geometric approach based on the concept of circuit complexity in quantum information theory. In a parallel vein, it has been proposed that certain deformations of the Euclidean path integral that prepares a given operator or state may provide an alternative definition, whose connection to the standard notion of complexity is less apparent. In this letter, we bridge the gap between these two proposals in two-dimensional conformal field theories, by explicitly showing how the latter approach from path integral optimization may be given by a concrete realization within the standard gate counting framework. In particular, we show that when the background geometry is deformed by a Weyl rescaling, a judicious gate counting allows one to recover the Liouville action as a particular choice within a more general class of cost functions.
\end{abstract}

\maketitle

\section{INTRODUCTION}

One of the most interesting developments of the past decade is the confluence of ideas from quantum gravity, quantum field theory (QFT), and quantum information science. The impact of this research trend is perhaps most striking in holography (AdS/CFT)~\cite{Maldacena:1997re,Witten:1998qj,Gubser:1998bc} where, e.g., the entanglement entropy of spatial subregions in the boundary field theory has been shown to play a prominent role in describing higher-dimensional (from this perspective, emergent) geometries; see ref.~\cite{Rangamani:2016dms} for a review. While the entanglement entropy is represented by extremal codimension-2 surfaces in the bulk~\cite{Ryu:2006bv,Hubeny:2007xt,Casini:2011kv,Lewkowycz:2013nqa,Dong:2016hjy}, it has been conjectured that certain codimension-1 \cite{Susskind:2014rva} and codimension-0 \cite{Brown:2015bva,Brown:2015lvg} bulk objects are related to the notion of ``complexity'' in the dual QFT. In particular, these objects appear to be natural probes of the black hole interior in AdS/CFT, the reconstruction of which is a timely problem of clear importance for our efforts to understand quantum gravity \cite{Freivogel:2014lja,Jefferson:2018ksk,Papadodimas:2013wnh,Papadodimas:2013jku,Papadodimas:2017qit,deBoer:2018ibj}.

These so-called ``holographic complexity'' proposals have not only contributed to ongoing research on the emergence of dynamical spacetimes from microscopic quantum mechanical degrees of freedom, but have also motivated the exploration of complexity in QFTs. The concept of complexity originates in quantum computing, and is based on approximating a given (typically unitary) operator in terms of some fundamental building blocks. In the context of quantum circuit design, the latter correspond to the gates used in constructing circuits that realize the given operation, the complexity of which is defined as the minimum-length circuit that achieves this goal. Note that, by fixing a reference state, this same framework allows one to speak of the complexity of states in the theory by optimizing over all circuits (operators) that produce the desired target state. From this perspective the complexity of operators is the more elementary notion, and it is this question on which we will focus in the present letter.

In practice, performing the optimization over all possible constructions is a difficult task. To surmount this problem in the original context of quantum circuit design, Nielsen and collaborators \cite{Nielsen:2005mn1,Nielsen:2006mn2,Nielsen:2007mn3} developed a procedure by which one could apply variational calculus to this optimization problem, by associating a geometry to the space of circuits based on the algebra of gates. Inspired by this approach, \cite{Jefferson:2017sdb} introduced the notion of circuit complexity in QFT by finding geodesics in the space of Gaussian states. Simultaneously, \cite{Chapman:2017rqy} put forth a complementary proposal based on continuous tensor network ideas \cite{Haegeman:2011uy,Nozaki:2012zj,Mollabashi:2013lya}. Together with subsequent developments \cite{Hackl:2018ptj,Khan:2018rzm,Camargo:2018eof,Chapman:2018hou,Guo:2018kzl,Bhattacharyya:2018bbv,Cotler:2018ufx,Balasubramanian:2018hsu}, these may be collectively considered as a geometric approach to complexity based on quantifying fundamental operations (i.e.\ gates). We shall refer to this line of thinking rather broadly as ``circuit complexity'', as per \cite{Jefferson:2017sdb}.

A second main approach based on so-called ``path integral optimization'' has been proposed in \cite{Miyaji:2016mxg,Caputa:2017urj,Caputa:2017yrh}. This was based on the observation that if one discretizes the Euclidean path integral which prepares a given state over a flat background, then a multiscale entanglement renormalization ansatz (MERA) \cite{MERA} can be loosely related to an optimization of this network which corresponds to performing the path integral over a Weyl-rescaled geometry. The resulting change in the path integral -- namely, the well-known Liouville action introduced below -- was then defined as the ``complexity'' of the corresponding state. This idea has since been made somewhat more precise in \cite{Milsted:2018yur,Milsted:2018san}; see also \cite{Czech:2017ryf} for a further development based on this prescription. However, in addition to certain technical subtleties, a conceptual drawback of this proposal is that it is not obvious why the Liouville action should be related to complexity in the usual sense of the word.

The aim of the present letter is to explicitly demonstrate that path integral optimization in its best-explored setting of two-dimensional conformal field theories (CFTs) may be given a precise formulation in terms of circuit complexity.

\section{REVIEW OF APPROACHES TO COMPLEXITY in QFT}

Let us begin by briefly reviewing the geometric approach introduced in refs.~\cite{Jefferson:2017sdb,Chapman:2017rqy}. In this approach, one represents the circuit -- i.e., the operator $V$ -- as a path-ordered exponential, consisting of a sequence of elementary ``gates'' generated by $\op_I$,
\small
\be
V = {\cal P} \exp{\left(- \int_{0}^{\lambda}\!\dd\kappa \, \sum_I Y^{I}\!(\kappa)\,\op_I \right)}~.
\label{eq:circuit}
\ee
\normalsize
In this expression, one can think of $\kappa$ as parametrizing the infinitesimal layers of the circuit, and $Y^{I}$ as controlling the gates inserted at that layer; in geometrical language, the latter may be regarded as a velocity component along the path $\kappa$. In order to measure the circuit length, one defines a cost function $\DD$, in which choice one has a considerable amount of freedom. For example, the choice most directly related to counting gates is the $L^1$-norm, $\DD_1 = \int_{0}^{\lambda}\!\dd\kappa \sum_{I}|Y^{I}|$. If one instead uses the $L^2$-norm $\DD_2 = \int_{0}^{\lambda}\dd\kappa \sqrt{\sum_{IJ}\eta_{IJ}Y^{I} Y^{J}}$ with $\eta_{IJ}$ some positive-definite matrix, then one arrives at the familiar problem of finding geodesics on a Riemannian manifold; see \cite{Jefferson:2017sdb} and related work above for details. Of course, there are other norms motivated from quantum information theory that serve equally well, in particular the Fubini-Study distance used in \cite{Chapman:2017rqy}.

We now turn to the path integral approach, in preparation to reframe the latter in the language of circuit complexity. We focus here on a specific example pioneered in ref.~\cite{Miyaji:2016mxg}, in which one works with an unnormalized thermal density matrix
\small
\be
\label{eq.rhobeta}
\rho_{\beta} = \exp{(-\beta H)}~,
\ee
\normalsize
where $H$ is the Hamiltonian operator of a two-dimensional CFT on a line. We regard $\rho_\beta$ as an operator $V$ (defined up to an overall normalization), rather than state, which we wish to decompose as in eq.~\eqref{eq:circuit}. This particular operator is of course very interesting since in addition to representing a thermal state in the CFT, in theories with a mass gap it also acts as a projector onto the vacuum state in the limit $\beta\rightarrow\infty$. 

\begin{figure}[h!]
\centering
\includegraphics[width=0.9\columnwidth]{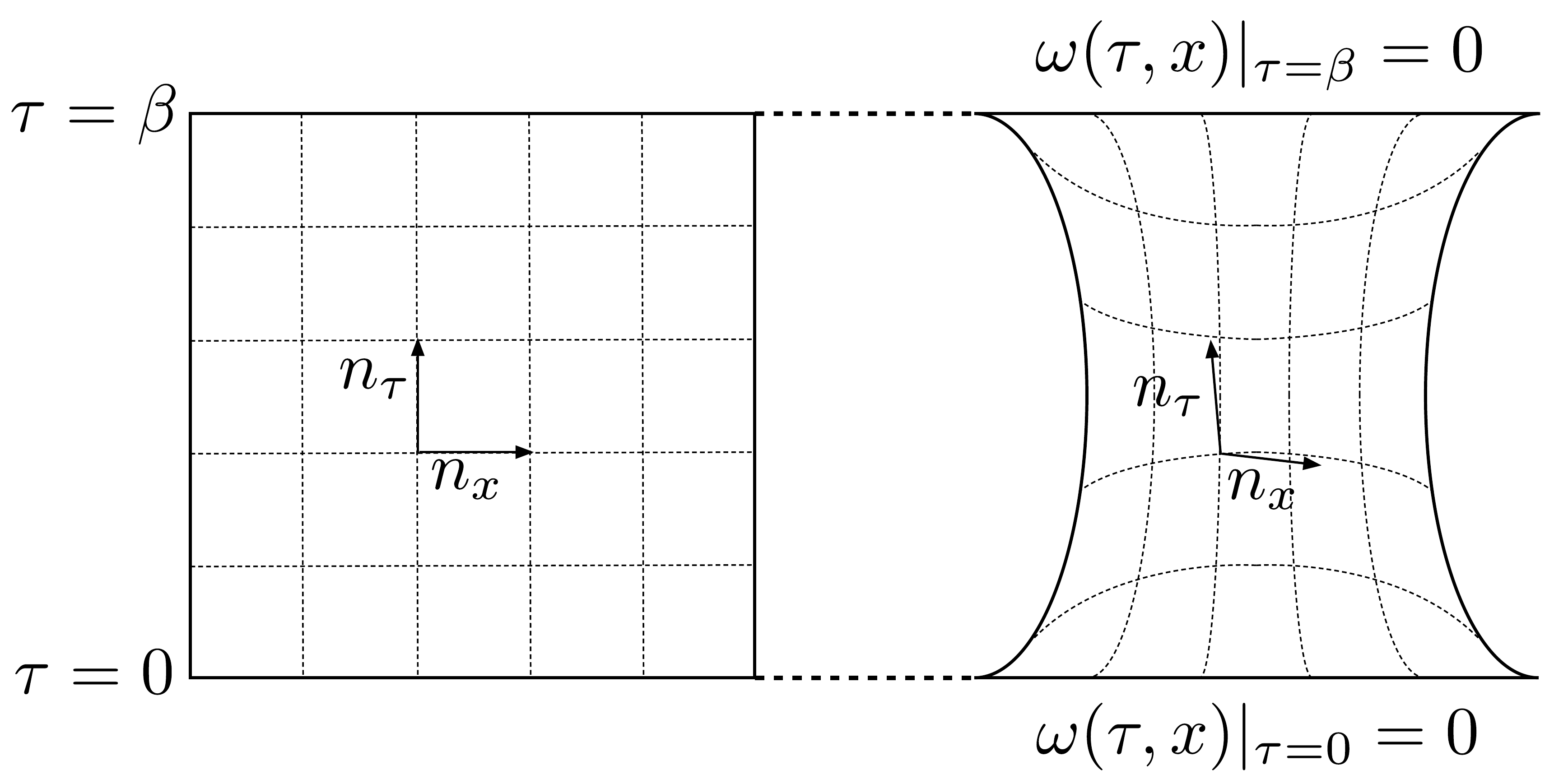}
\vspace{-5pt}
  \caption{Euclidean strip over which we perform the path integral \eqref{eq:A}. The left image represents the flat geometry \eqref{eq:flat}, while the right represents the Weyl-rescaled background \eqref{eq.Weylresc}. Note that we hold the boundary conditions at $\tau=0$ and $\tau=\beta$ fixed. This guarantees that we produce the same operator $\rho_{\beta}$, up to an overall normalization governed by the exponent of the Liouville action~\eqref{eq:SL}.\label{fig:Weylresc}}
\end{figure}

The matrix elements of $\rho_\beta$ may be represented via a Euclidean path integral over a strip of flat geometry,
\small
\be
\dd s_{0}^2 = \dd\tau^2 + \dd x^2~,\label{eq:flat}
\ee
\normalsize
with width $\beta$ in the $\tau$-direction:
\small
\be
\bra{\phi(\beta,x)}\rho_{\beta}\ket{\phi(0,x)}
=\int_{\hat\phi(0,x)=\phi(0,x)}^{\hat\phi(\beta,x)=\phi(\beta,x)}\DD\hat\phi\,e^{-S_{\mathrm{CFT}}[\hat\phi]}~,\label{eq:A}
\ee
\normalsize
where $\hat\phi$ denotes the fields in the theory and $\phi$ particular eigenstates of field operators. It then transpires that if one deforms the background geometry over which the path integral is performed by a Weyl factor $\omega(\tau,x)$,
\small
\be
\label{eq.Weylresc}
\dd s_{\omega}^2 = g_{\mu\nu}\dd x^{\mu}\dd x^{\nu} = e^{2\, \omega(\tau,x)} \left(\dd\tau^2 + \dd x^2 \right)~,
\ee
\normalsize
such that $\omega(\tau\!=\!0,x)=\omega(\tau\!=\!\beta,x)=0$, then the same operator $V$ is prepared (again up to an overall normalization), cf.\ fig.~\ref{fig:Weylresc}. 
As alluded in the introduction, refs.~\cite{Caputa:2017urj,Caputa:2017yrh} then proposed that a natural cost function to consider is the Liouville action
\small
\be
S_L=\frac{c}{24\pi}\int_{0}^\beta\!\dd\tau\int_{-\infty}^{\infty}\!\dd x\left[\Lambda \, e^{2\omega} + \dot\omega^2 + \omega^{\prime 2}\right] ,\label{eq:SL}
\ee
\normalsize
where $\dot{\phantom{a}}$ and $'$ denote derivatives w.r.t.\ $\tau$ and $x$, respectively, and $c$ is the central charge. The Liouville action governs the change in the path integral measure under the above Weyl rescaling. Here, $\Lambda\sim\epsilon^{-2}$, where $\epsilon$ denotes some UV cutoff in real space (i.e., the lattice spacing).

In the present letter we will demonstrate that, up to an unimportant overall normalization, $S_{L}$ corresponds in a precise way to a particular cost function in the geometric approach to circuit complexity outlined above. Before proceeding, let us stress two aspects of the Liouville action that will be relevant later. 
The first is that the Liouville action is properly a covariant expression, which in arbitrary coordinates~$\xi^{\mu}$ can be shown to take the form \cite{Polyakov:1981rd}
\small
\be
\label{eq:SLcov}
S_L=\frac{c}{24\pi} \int\!\dd^2 \xi \sqrt{g} \left[ \Lambda + \frac{1}{4}\partial_{\mu} \chi \partial^{\mu} \chi \right],
\ee
\normalsize
where the scalar $\chi$ is a covariant albeit non-local expression defined in terms of the Ricci scalar~$R$ and the propagator $\Box^{-1}$ for a massless scalar field,
\small
\be
\label{eq.chidef}
\chi(\xi) = \int\!\dd^2 \tilde{\xi} \sqrt{g(\tilde{\xi})} \, \Box^{-1}(\xi,\tilde{\xi}) R(\tilde{\xi}).
\ee
\normalsize
{\color{change}Furthermore, we impose that $\chi$ vanishes at the boundaries to make it unique.}
Note that by changing the relative normalization between the $\dot\omega^2$ and $\omega^{\prime2}$ terms in eq.~\eqref{eq:SL}, one would not be able to recover the covariant form~\eqref{eq:SLcov}. The second comment is that one should regard the Liouville action as consisting of the two leading terms in the expansion in derivatives of the metric weighted with respect to $\epsilon$: the cosmological constant term $\Lambda$, which is ${\cal O} \left(\epsilon^{-2}\right)$ and diverges in the continuum, is related to the energy-momentum tensor renormalization, while the nonlocal term is ${\cal O} \left(\epsilon^{0}\right)$ and gives rise to the trace anomaly. One could add terms having more derivatives, but they would necessarily come with positive powers of~$\epsilon$ and hence vanish in the $\epsilon\rightarrow0$ limit. While these terms might become important for the path integral optimization program, as we discuss below, we ignore them in the present letter.

\section{EUCLIDEAN PATH INTEGRALS AS CIRCUITS}

We now wish to interpret the Euclidean path integral on the Weyl-rescaled geometry as a circuit \`a la \eqref{eq:circuit}. To do so, we will make use of a formula developed in refs.~\cite{Milsted:2018yur,Milsted:2018san} in the context of tensor networks linking path integrals on curved geometries with the exponentiation of the energy-momentum tensor components of two-dimensional CFTs in Minkowski space. 

\begin{figure}[h!]
\centering
\includegraphics[width=0.9\columnwidth]{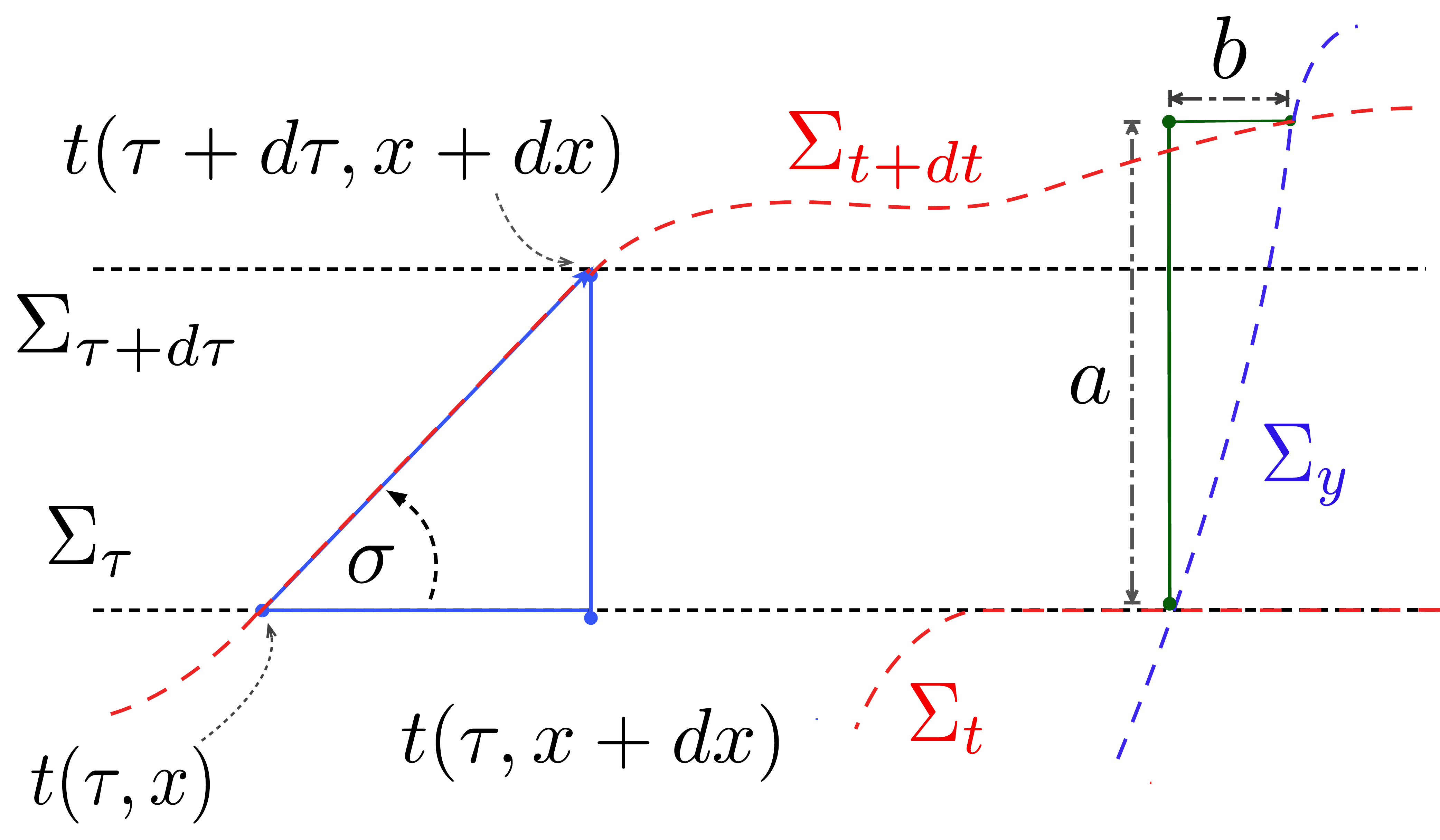}
\vspace{-5pt}
\caption{$\Sigma_\tau$ and $\Sigma_t$ denote time slices of the background geometry in $(\tau,x)$ coordinates, see eq.~\eqref{eq.Weylresc}, and $(t,y)$ coordinates, see eq.~\eqref{eq:gMV}, respectively. At a given point, the Euclidean angle between tangents to $\Sigma_{t}$ and $\Sigma_{\tau}$ is given by the function $\sigma(\tau,x)$, as defined in eq.~\eqref{eq.defsigma}. On the right side of the figure, $a$ and $b$ appear as defined by eq.~\eqref{eq:gMV}, which illustrates their geometrical interpretation.}
\label{fig.fromPItoOs}
\end{figure}

The relevant statement is the following. If one considers a Euclidean path integral for a CFT$_{2}$ on a generic background in coordinates in which constant Euclidean time $t$ slices are flat lines, i.e.,
\small
\be
\label{eq:gMV}
\dd s^2=\left(a(t,y)^2 + b(t,y)^2\right) \,\dd t^2+2\,b(t,y)\,\dd t\,\dd y+\dd y^2~,
\ee
\normalsize
then, up to an unimportant normalization, this path integral computes the matrix elements of the operator
\small
\be
\label{eq.Tmunucircuit}
V = \PP\exp\left\{-\int_{t_{i}}^{t_{f}}\!\dd t \int\!\dd y \,\left[ a(t,y)\,h(y)+i\,b(t,y)\,p(y)\right]\right\}~,
\ee
\normalsize
where $h\equiv T_{t_{M}t_{M}}$ and $p\equiv T_{t_{M}y}$ are given in terms of the stress tensor in Minkowski space, whose line element we write as $-\dd t_{M}^2 + \dd y^2$, and the path ordering is applied to the integration over the Euclidean time $t$. One can understand this expression pictorially in a very simple way, as illustrated in fig.~\ref{fig.fromPItoOs}.

We then observe that eq.~\eqref{eq.Tmunucircuit} has the same form as our circuit \eqref{eq:circuit}. Accordingly, we interpret the path integral \eqref{eq.Tmunucircuit} as a bilinear map that acts on the Hilbert space on a slice of Euclidean time $t$.  Comparing with eq.~\eqref{eq:circuit}, the circuit parameter $\kappa$ is then identified with $t$, and the generators of the gates $\op^{I}$ are the energy-momentum components on a slice of flat Minkowski space, namely the Hermitian and anti-Hermitian operators $h$ and $ip$, respectively. We do not include the pressure $T_{yy}$ since the tracelessness of the energy-momentum tensor makes it equal, as an operator, to the energy density $h$. The index $I$ runs over the full range of $y$ and differentiates between $h$ and $i \, p$. Note that in contrast to the existing literature on circuit complexity reviewed above, we are working with circuits consisting of both Hermitian (Euclidean) and unitary transformations. This is a natural consequence of the fact that the Wick rotation to Euclidean time places the $tt$ and $ty$ components of the stress tensor on different footing.

To proceed, we must identify coordinate transformations from the $(\tau,x)$ coordinates in the Weyl-rescaled form of the metric \eqref{eq.Weylresc} to the $(t,y)$ coordinates \eqref{eq:gMV}. 
The most general coordinate transformation can be parametrized by a diffeomorphism $t(\tau,x)$, $y(\tau,x)$. 
{\color{change} In appendix A we develop some intuition for the particular case where only $y=y(\tau,x)$ is transformed.}
One can then solve for $a$ and $b$ in terms of $t$ and $y$, which leads to the following result:
\small
\be
\label{eq.abinterestingcircuits}
a = \frac{\dot t \, y^\prime - t^\prime \, \dot y}{\dot t^2 + t^{\prime 2}} 
\qquad \mathrm{and} \qquad
b = -\frac{\dot t \, \dot y + t^\prime \, y^\prime}{\dot t^2 + t^{\prime 2}}~.
\ee
\normalsize
Of course, one also needs to ensure that $y(0,x) = y(\beta,x) = x$, and that $t(0,x)$ and $t(\beta,x)$ are constant, as per the boundary conditions explained above. 
In this most general case we have the freedom to choose two functions subject to the aforementioned boundary conditions. It is important to note that, while we think of $a$ and $b$ as functions specifying the circuit, a generic choice of these functions will not return a circuit of the form \eqref{eq.Tmunucircuit} that represents, up to normalization, the operator $\rho_{\beta}$~\eqref{eq.rhobeta}. However, the specific parametrization~\eqref{eq.abinterestingcircuits} guarantees this. 

Before we proceed to specifying the cost function, it will be very convenient to introduce an angle $\sigma$, defined~as
\small
\be
\label{eq.defsigma}
{t'} = -  {\dot{t}} \, \tan{\sigma}~,
\ee
\normalsize
the intuition for which is explained in fig.~\ref{fig.fromPItoOs}. Furthermore, we note that derivatives of $t$ and $y$ are related to the Weyl factor $\omega$ by
\small
\be
y^\prime = e^\omega \sec\sigma - \dot y \tan\sigma~.
\ee
\normalsize
In what follows, we will count gates in terms of cost functionals of $a,\,b$ \eqref{eq.abinterestingcircuits}; since our goal is to reproduce the Liouville action naturally expressed in terms of the Weyl parameter, we will aim to reexpress these cost functionals in terms of $\omega$ and the free (gauge) parameter $\sigma$.

\section{COST FUNCTIONS AND THE LIOUVILLE ACTION\label{sec:costfcts}}

In counting gates, our point of departure is regarding eq.~\eqref{eq.Tmunucircuit} as a circuit of the form~\eqref{eq:circuit}. Of course, as in previous works on the geometric approach to circuit complexity, there are many cost functions one could consider. For example, a natural choice is an $L^1$-norm; while this is ultimately not the choice we wish to make, it is helpful for gaining some intuition about the cost function, {\color{change}and we refer the reader to the appendix~B for a discussion of this case and appendix C for the $\kappa=2$-norm as another common, albeit problematic choice of the cost function.} A further choice is the more familiar $L^2$-norm; but this is rather more subtle than the $L^1$-norm, since this puts the integral over $y$ under the square root, and hence does not lead to natural expressions in terms of $\omega$ and $\sigma$ as functions of $\tau$ and $x$. Furthermore, if we wish to obtain the Liouville action as a proper cost function, it appears that the most appropriate choice lies between these two familiar norms, and takes the form
\small
\be
\label{eq.Lalt}
\DD_{\mathrm{L}} \simeq \int\!\dd t\, \dd y \,\frac{1}{\epsilon^2}\, \sqrt{a^2 + \epsilon^2 \eta_{(\partial a)^2} (\partial_{y} a)^2 + \epsilon^2 \, \eta_{(\partial b)^2} (\partial_{y} b)^2 + \ldots}~,
\ee
\normalsize
where the ellipsis again denotes higher derivative terms which we shall drop, and we have neglected more complicated terms of the sort discussed above. As alluded above, treating this expression exactly does not result in a diffeomorphism-invariant (i.e., $\sigma$-independent) cost function. However, if we choose equal penalty factors, i.e.~$\eta_{(\partial a)^2} = \eta_{(\partial b)^2}$, and expand to next-to-leading order in $\epsilon$, we find that $\omega$ and $\sigma$ decouple, and our cost function takes the form
\small
\be
\ba
\DD_\mathrm{L} \simeq&\int\!\dd\tau\, \dd x \, \bigg\{ \frac{e^{2 \omega}}{\epsilon^2} + \frac{1}{2}\eta_{(\partial a)^2} \left(\dot{\omega}^2 + \omega'^2 \right)\\
&+\frac{1}{2} \eta_{(\partial a)^2} \left( \dot{\sigma}^2 + \sigma'^2 \right) + \eta_{(\partial a)^2} ( \omega' \dot{\sigma} - \dot{\omega} \sigma') + \ldots\bigg\}~.
\ea\label{eq.LaltLiouville}
\ee
\normalsize
Observe that the last term to this order is a total derivative, and hence the optimization over $\omega$ is independent from the choice of $\sigma$. Furthermore, the $\omega$-dependent part is none other than the Liouville action. This implies that the Liouville action that underlies the approach to complexity from ``path integral optimization'' can be given a precise interpretation as an approximation to a particular choice of cost function. However, from this perspective there appears nothing fundamental about the Liouville action as a measure of complexity. Indeed, the former simply arises from the change in the path integral measure under Weyl rescaling, as reviewed above, and we have relied on a degree of fine-tuning of the cost function in order to recover a complexity measure of the same form. A similar observation was made in \cite{Takayanagi:2018pml}, where the Liouville action was found to be the next-to-leading order term in an $\epsilon$-expansion of a bulk gravitational action with a boundary term in AdS$_3$ with a position dependent cutoff.

As a closing remark, we note that for $\sigma = 0$ the cost function~\eqref{eq.Lalt} resembles a Dirac-Born-Infeld (DBI) action (for a review, see \cite{Johnson:2003gi}) for the field $\omega$, and hence in terms of the field $\chi$ defined in eq.~\eqref{eq.chidef}, we may write
\small
\be
\label{eq.LaltDBI}
\mathcal D_{\text{DBI}} \simeq \int\!\dd^2\chi \sqrt{\mathrm{det}\left( g_{\mu \nu} + \epsilon^2 \partial_{\mu} \chi \partial_{\nu} \chi \right)}~.
\ee
\normalsize
It would be interesting to pursue this analogy further, but we leave this for future work.

\section{DISCUSSION AND OUTLOOK}

This letter can be regarded as an attempt to define circuit complexity in QFT as a direct functional of sources appearing in the underlying Euclidean path integral. Furthermore, as opposed to the earlier works on gate complexity in QFT, we explicitly use only local gates. In this framework, we were able to show that the cost function appearing in the path integral optimization approach, namely the Liouville action given by eq.~\eqref{eq:SLcov}, can be thought of as an approximation to a genuine gate counting procedure in which the source is the background metric. 

One aspect to emphasize is that the covariance with respect to an underlying metric is lost for a genuine cost function, since these are defined with respect to some time foliation. So far we were only able to recover covariance approximately to next-to-leading order in the UV cutoff in the relevant expansion for obtaining the Liouville action, cf.~eq.~\eqref{eq.LaltLiouville}. In general, one expects that covariance combined with a proper gate counting prescription poses quite restrictive constraints on allowed metric functionals. 
On this front, note that in the path integral optimization approach, one just varies the cost function with respect to $\omega$. However, in the covariant approach one would also vary with respect to other metric components, and this should lead to some constraint equations. 

Another important item that we want to stress in the context of Liouville is that our cost functions are not sensitive to the central charge of the CFT. Of course, one has the freedom to introduce the central charge in the overall normalization (or in penalty factors), but the arguments above do not provide any physical justification for doing so, and consequently we have only considered cost functions defined up to some overall prefactor.

There are two further implications of this letter we wish to highlight. First, a proof of the holographic complexity proposals should occur through the equality of bulk and boundary Euclidean partition functions, and the interface between the two sides is governed by the boundary (QFT) sources. While there have already been several attempts in this general direction \cite{MIyaji:2015mia,Belin:2018fxe,Belin:2018bpg,Bernamonti:2019zyy}, the present paper is perhaps the first to construct genuine cost functions for a particular class of source configurations. Second, in the path integral optimization program \cite{Miyaji:2016mxg,Caputa:2017urj,Caputa:2017yrh}, one optimizes the Liouville action alone, i.e., without including higher order corrections in $\epsilon$. Since we found it rather non-trivial to construct a cost function that formally matches Liouville, while the seemingly more natural ones included higher order corrections in $\epsilon$, one should keep in mind that perhaps the optimal circuits / geometries for genuine cost functions might be different than the ones obtained using the Liouville action alone. On the other hand, the latter were found to correspond to geometries consistent with hyperbolic spacetimes; in light of the complexity=volume proposal, one may therefore wish to keep this feature as an additional condition beyond covariance and gate counting. In this spirit, eq.~\eqref{eq.LaltDBI} might be taken as an interesting point of departure for studies of cost functions to all orders in~$\epsilon$.

Regarding direct extensions of this work, it would be very interesting to find a cost function that is fully covariant to all orders in $\epsilon$. Furthermore, an approach closely related to the use of the Fubini-Study distance has been developed for unitary transformations with energy-momentum tensor insertions in \cite{Caputa:2018kdj}, building on earlier results in~\cite{Magan:2018nmu}. Although \cite{Caputa:2018kdj} focused on purely holomorphic or antiholomorphic transformations, it seems important to understand the relation with the present work in greater detail. One attempt in this direction might be to use our circuits contracted with some sample state (for example, the family of states used in \cite{Caputa:2018kdj}) and in this way calculate the Fubini-Study distance along the lines of ref.~\cite{Chapman:2017rqy}. Furthermore, when compensating for the derivative dimensions in the present paper, we relied on the UV regulator $\epsilon$ as the only obvious scale in the problem. However, if one has insertions of operators at different spatial positions on a constant $t$ slice, say at $y$ and $\tilde{y}$, then the Euclidean distance between them acts as another, relational dimensionful quantity. 
Additionally, the current approach may be generalized to provide a definition of complexity for mixed states by including non-unitary transformations; cf.~\cite{Agon:2018zso,Camargo:2018eof} for earlier works in this vein.

\begin{acknowledgments}
\begin{center}\textbf{ACKNOWLEDGEMENTS}\end{center}

We wish to thank E.~Brehm, P.~Caputa, B.~Czech, M.~Flory, B.~Freivogel, L.~Hackl, D.~Hofman, L.~Hadasz, R.~Janik, J.-L.~Lehners, T.~Osborne, V.~Schomerus, V.~Svensson, S.~Theisen, T.~Takayanagi, and H.~Wang for helpful discussions and correspondence and, especially, J.~Hung for her key insight inspired by DBI actions that allowed us to derive the Liouville action as an approximation to a cost function. We are also very grateful to D.~Das and S.~He for their input during earlier stages of the project. Our special thanks go also to S.~Singh, M.~Walter, and F. Witteveen, as well as J.~de~Boer, S.~Chapman, and I.~Reyes for their feedback and collaboration with some of us on related topics. The Gravity, Quantum Fields and Information group at AEI is generously supported by the Alexander von Humboldt Foundation and the Federal Ministry for Education and Research through the Sofja Kovalevskaja Award. H.~C.\ is partially supported by the Konrad-Adenauer-Stiftung through their Sponsorship Program for Foreign Students.
\end{acknowledgments}

\bibliographystyle{jk_ref_layout_noTitle} 
\bibliography{biblio}

\clearpage
\appendix*
\onecolumngrid
\setcounter{equation}{0}

\begin{center}\textbf{PATH INTEGRAL OPTIMIZATION AS CIRCUIT COMPLEXITY:\\
SUPPLEMENTARY MATERIAL (APPENDIX)}\end{center}

In this supplementary material we discuss some further details of our derivation of path integral optimization from the circuit complexity framework.
In appendix A some remarks on an intuitive understanding of the coordinate transformations are presented.
In appendix B and C we elaborate on two other important norms that allow us to define cost functions and discuss their properties.

\section{A.~   Intuitive understanding of coordinate transformations}

The simplest coordinate transformation of the kind considered in the main text is
\be
\label{eq.simpletrafo}
t = \tau \qquad \mathrm{and} \qquad y = \int_{0}^{x}\!\dd\gamma\,e^{\omega(\tau,\gamma)}~.
\ee
One can easily convince oneself that this transformation brings the Weyl-rescaled metric \eqref{eq.Weylresc} to the metric ansatz \eqref{eq:gMV}. Furthermore, it yields the following expressions for the metric components $a$ and $b$:
\be
a = y' = e^{\omega(\tau,x)},
\;\;\;
b = -\dot{y} = -\int_{0}^{x}\!\dd\gamma\,e^{\omega(\tau,\gamma)}\dot\omega(\tau,\gamma)~.
\label{eq.absimpletrafo}
\ee
We may now rewrite eq.~\eqref{eq.Tmunucircuit} explicitly in terms of these components; note that converting to $(\tau,x)$ coordinates introduces a factor of $e^{2\omega}$ from the Jacobian, whereupon we have
\be
V = \PP\exp\,\Bigg\{ -\int_{0}^{\beta}\!\dd \tau \int_{-\infty}^{\infty}\!\dd x \,\bigg[h(x)-
\lp e^{-\omega(\tau,x)} \int_{0}^{x}\!\dd\gamma\,\dot\omega(\tau,\gamma)e^{\omega(\tau,\gamma)}\rp i\,p(x)\bigg]\Bigg\} .
\label{eq.Weylcircuitintuit}
\ee
Let us now consider this expression from the point of view of the linear map generated by the path integral on the metric~\eqref{eq.Weylresc} from some time slice $\tau$ to a time slice $\tau + \delta \tau$. The first term in the exponent~\eqref{eq.Weylcircuitintuit} generates the relevant time translation. However, as we move from one slice to the other, the geometry in the transverse direction also changes. To account for this, we need precisely the second term in the exponent, which in transitioning between time slices acts as an infinitesimal, position-dependent translation $x \rightarrow x + u(x)\,\delta \tau$. Indeed, equating the change in the transverse direction with the action of this translation, we have
\be
e^{2\,\omega(\tau + \delta \tau,x)} \, \dd x^2 = e^{2\omega(\tau,x + u(x)\, \delta\tau)} \, \dd\lp x + u(x)\, \delta\tau\rp^2~,
\ee
whereupon expanding to first order in $\delta\tau$ and solving for $u(x)$ returns precisely the term multiplying $i \, p(x)$ in eq.~\eqref{eq.Weylcircuitintuit}.

\section{B.~   Properties of \boldmath{$L^{1}$} cost functions}
\label{app:Intuition}

In order to develop some intuition about the cost function used in this letter, let us consider the $L^1$-norm mentioned in the main text. In the simplest case, a direct counting of $a$ and $b$ in the circuit would lead to
\be
\label{eq.L1costsimp}
\DD_{1} \simeq \int\!\dd t\, \dd y \,\left(|a| + \eta_{b} |b| \right)~,
\ee
where the overall normalization does not matter, and we always associate some cost to progressing in the time direction (i.e., along the circuit). We may also allow for a different cost associated with the transverse direction, hence the penalty factor $\eta_{b} \geq 0$ (see \cite{Jefferson:2017sdb} for a discussion of penalty factors in this context). 
Given eq.~\eqref{eq.absimpletrafo} however, we immediately see that this cost function will be a rather complicated functional of $\omega$, even for the simplest case $\sigma=0$. To avoid this, we choose $\eta_b=0$. Physically, this corresponds to making uniform space translations free.

In general, there is little reason to stop at zeroth-order derivatives in \eqref{eq.L1costsimp}. However, in order to make contact with the definition of circuit complexity reviewed in this letter, we exclude derivatives with respect to $t$, since these would correspond to associating a cost to ``acceleration'' along the path. Additionally, since derivative terms are dimensionful, they must be weighted by some UV regulator $\epsilon$, cf. eq.~\eqref{eq:SL}. For the same reason, we must include an overall factor of $\epsilon^{-2}$ from the integration measure, whereupon a more sophisticated version of \eqref{eq.L1costsimp} is
\be
\label{eq.L1costcomplicated}
\DD_{1} \simeq \int\!\dd t\, \dd y \,\frac{1}{\epsilon^2}\, \left(|a| + \epsilon \, \eta_{\pd a} |\partial_{y} a| + \epsilon \, \eta_{\pd b} |\partial_{y} b| + \ldots \right)~,
\ee
where we have included possible penalty factors $\eta_{\pd a}$, $\eta_{\pd b}$, and the ellipsis denotes terms ${\cal O}(\epsilon^{0})$. Relative to the simpler expression \eqref{eq.L1costsimp}, we are now accounting for the cost of applying $h$ uniformly, we do not care about uniform applications of $p$, and, more importantly, there is a cost associated with moving in $t$ or $y$ when neighbouring points on a constant $t$-slice are transformed non-uniformly. We must note however that even to this order in the formal $\epsilon$ expansion, this expression is not the most general one could consider: derivative terms imply a gate counting with respect to some basis that couples neighboring sites, and hence terms such as $|\partial_{y} a + \partial_{y} b |$ would be equally valid. 

The final step is to express this cost function in $(\tau,x)$ coordinates, which introduces a Jacobian. Despite the complicated dependence on $t$ and $y$ and their derivatives, the result admits a relatively simple expression in terms of $\omega$ and $\sigma$:
\be
\DD_1 \simeq \int_{0}^{\beta}\!\dd \tau \int_{-\infty}^{\infty}\!\dd x \, \frac{e^{\omega}}{\epsilon^2} \bigg\{e^{\omega}
+\epsilon\,\eta_{\partial a}\big| (\dot{\omega} - \sigma') \, \sin{\sigma} +  (\omega' + \dot{\sigma}) \, \cos{\sigma} \big| 
+\epsilon\,\eta_{\partial a}\big| (\omega' + \dot{\sigma}) \, \sin{\sigma} - (\dot{\omega} - \sigma') \, \cos{\sigma}
 \big|~.
\label{eq.L1costexplicit}
\ee
One crucial aspect that we wish to highlight is that since this cost function depends explicitly on the foliation into constant time slices via $\sigma$ (cf.~fig.~\ref{fig.fromPItoOs}), it is not a diffeomorphism-invariant cost function of the metric $g_{\mu \nu}$, and indeed this property is not manifest in our approach. Indeed, even if we consider the flat space case, i.e. $\omega = 0$, our cost function would still be non-trivial and foliation-dependent.

To close the discussion of the $L^{1}$ cost function, let us consider the special case in which the Weyl factor is translation-invariant, $\omega(\tau)$, whereupon $\sigma=0$ is the most natural condition consistent with the boundary conditions at $\tau=0$ and $\tau = \beta$. This leads to
\be
\label{eq.L1costexplicitexample}
\DD_{1}^{\mathrm{ex}} \simeq \int_{0}^{\beta}\!\dd \tau \, \frac{1}{\epsilon} \left\{
e^{2\,\omega} + \epsilon \, \eta_{\partial B} |e^{\omega}\, \dot{\omega}| \right\}~,
\ee
where we dropped higher order terms and neglected an unimportant prefactor.
We now ask what the minimum of this cost function looks like. Recall that we always impose $\omega=0$ at the boundaries. In the most general situation, the first term takes the minimal value for $\omega \rightarrow -\infty$. Normally in physical problems it is the kinetic term that penalizes such run-away behaviour, but in this case the kinetic term is an absolute value of a total derivative $\frac{\dd}{\dd\tau} e^{\omega}$. As a result, it does not care about how fast the function changes, but exhibits only a piecewise sensitivity to the total change. This implies that the optimal circuit includes an almost infinite dilatation, followed by Euclidean time evolution with an extremely coarse-grained Hamiltonian, and then another almost infinite dilatation to satisfy the boundary condition at $\tau=\beta$. This situation bears a striking resemblance to tensor network renormalization \cite{2015PhRvL.115r0405E}, with the identification of MERA \cite{Vidal:2007hda} as the almost infinite dilations, and the Euclidean time evolution in between as evolution under a single layer of so-called ``Euclideons'' \cite{2015PhRvL.115t0401E}.

\section{C.~   Properties of \boldmath{$\kappa=2$} cost functions}

What is often used in geometric gate counting is the-so called $\kappa=2$-function (strictly speaking, it is not really a norm). It comprises a quadratic counting of gates, given by
\be
\label{eq:Dkappa2}
\DD_{\kappa=2} \simeq \int\!\dd t\, \dd y \,\frac{1}{\epsilon^2}\, a^2 + \eta_{(\partial a)^2} (\partial_{y} a)^2 + \eta_{(\partial b)^2} (\partial_{y} b)^2 + \ldots~.
\ee
Similarly to the previous expressions, $\eta_{(\partial a)^2}$ and $\eta_{(\partial y)^2}$ are penalty factors. 
Higher order derivatives would appear with a positive power of the regulator and vanish in the UV limit $\epsilon \to 0$, while mixed terms such as $\frac{1}{\epsilon}a\partial_y B$, $\partial_y a \partial_y B$ are in principle allowed but neglected here.
Expressed in terms of velocity components, the $\kappa=2$-norm can be written as
\be
\DD_{\kappa=2} \simeq \int\!\dd t \sum_{IJ} \eta_{IJ} Y^I Y^J~,
\ee
where the metric $\eta_{IJ}$ takes the form
\be
\eta_{IJ} = 
\begin{pmatrix}
(-\partial_y^2 + \mu^2) \delta(y-y^\prime) & 0 \\
0 & -\partial_y^2 \delta(y-y^\prime)
\end{pmatrix}~,
\ee
such that the gate counting can be re-expressed as
\be
\DD_{\kappa=2} \simeq \int\!\dd t \int\!\dd y \int\!\dd y^\prime \ \begin{pmatrix} a(y) & B(y) \end{pmatrix} \eta_{IJ} \begin{pmatrix} a(y^\prime) \\ B(y^\prime) \end{pmatrix} .
\ee
(The sum $\sum_{IJ}$ has been rewritten as a double integral over the label $y$ and a summation over the gates $a$, $B$; penalty factors are suppressed.)
Transforming the cost function into the $(\tau,x)$ coordinate system, one encounters a Jacobian $J = \dot t y^\prime - t^\prime \dot y = e^\omega \sec(\sigma) \dot t$, where the definition of $\sigma$ in \eqref{eq.defsigma} is employed.
The cost function then reads
\be
\label{eq:Dkappa2_taux}
\DD_{\kappa=2} \simeq \int_{0}^{\beta}\!\dd \tau \int_{-\infty}^{\infty}\!\dd x \, \frac{e^\omega \cos\sigma}{\dot t} \left\{  \frac{e^{2\omega}}{\epsilon^2} + \eta_{(\partial a)^2}(\sigma^\prime-\dot\omega)^2 + \eta_{(\partial a)^2}(\dot\sigma+\omega^\prime)^2 \right\} 
\ee
for equal penalty factors $\eta_{(\partial a)^2} = \eta_{(\partial b)^2}$.
The crucial property of this cost function is the fact that this expression depends explicitly on the choice of $t(\tau,x)$, i.e.\ the circuit parametrization, via the derivative $\dot t$ appearing in the denominator.
Contrary to the gate counting \eqref{eq.Lalt}, which resembles a Schatten norm and does not depend on $t(\tau,x)$, or the $L^1$-norm \eqref{eq.L1costexplicit}, one now has $\omega$, $\sigma$ and $t$ as variational parameters.
In the most general case, the resulting equations of motion for the minimization are complicated and do not decouple.
Demanding agreement with the Liouville action in every power of $\epsilon$ does not give consistent conditions for the choice of the time slicing $t(\tau,x)$.
(We neglect intricate subtleties such as total derivatives in comparing the cost function \eqref{eq:Dkappa2_taux} with the Liouville action.)
An exception for this statement is the special case of a translational invariant circuit $\omega = \omega(\tau)$.
Under this condition, one can set $\sigma=0$ and choose the particular time slicing $\dot t = e^{\omega(\tau)}$.
This cancels the prefactor appearing in \eqref{eq:Dkappa2_taux} and one recovers the Liouville action.

\end{document}